\documentclass[prd,aps,floats,twocolumn]{revtex4}
\usepackage{epsfig}

\newcommand{\be}{\begin{equation}}
\newcommand{\ee}{\end{equation}}
\newcommand{\ba}{\begin{eqnarray}}
\newcommand{\ea}{\end{eqnarray}}
\newcommand{\bc}{}

\begin{document}

\preprint{\
\begin{tabular}{rr}
&
\end{tabular}
}
\title{The Cosmology of a Universe with Spontaneously-Broken Lorentz Symmetry}
\author{P.G.~Ferreira$^{1,2}$,
B.M.~Gripaios$^{1}$,
R.~Saffari$^{1,3}$,
T.G.~Zlosnik$^{1}$}
%
\affiliation{
$^{1}$Astrophysics and Theoretical Physics, University of Oxford, Denys Wilkinson Building, Keble Road, Oxford OX1 3RH, UK\\
$^{2}$African Institute for Mathematical Sciences, 6-8 Melrose Road, Muizenberg 7945, South Africa\\
$^{3}$Institute for Advanced Studies in Basic Sciences,\\ Gava
Zang, P.O.Box 45195-1159 Zanjan, Iran }

\begin{abstract}
A self consistent effective
field theory of modified gravity has recently been proposed with spontaneous breaking of local
Lorentz invariance. The symmetry is broken by a vector field with the
wrong-sign mass term and it has been shown to have additional graviton modes
and modified dispersion relations. In this paper we study the evolution of
a homogeneous and isotropic universe in the presence of such a vector field
with a minimum lying along the time-like direction. A plethora of different
regimes is identified, such as accelerated expansion, loitering, collapse and
tracking.
\end{abstract}

\date{\today}
\pacs{PACS Numbers : }
\maketitle

\noindent
\section{Introduction}
The evolution and current state of the universe has become amenable
to a number of astronomical and experimental probes. In the past
few years, it has become possible to measure its geometry, expansion
rate and constituents with unprecedented precision. The recent
measurement of the anisotropy and polarization of the cosmic
microwave background with the Wilkinson Anisotropy Microwave Probe satellite
has further improved constraints on cosmological parameters \cite{wmap3}.
We now believe that we have identified an accurate and consistent
model of the universe based on the general theory of relativity.

However our current model relies on the existence of extremely exotic
forms of energy: dark matter which clumps gravitationally but does
not interact with light and dark energy which is repulsive under gravity.
Furthermore, we must posit that $95\%$ of the energy density of
the universe is taken up by these exotic components. There are
proposals for their fundamental origin but as yet no compelling
explanation.

An alternative possibility is that gravity is not what it seems and that
general relativity should be modified on certain scales.
A notable example of this is the relativistic theory of Modified Newtonian
Dynamics \cite{milgrom, teves}. In this theory, dark matter is replaced by a modification
of the gravitational interaction, through the inclusion of a scalar and
vector field. These extra fields can compensate for the absence of
cold dark matter on galactic and galactic-cluster scales, and at the cosmological horizon.
Other proposals involving additional metrics and a combination of scalar,
vector and tensor fields have been proposed. Typically these models are
constructed with the goal of reproducing observation but with little or no
basis on fundamental principles. Mostly they are plagued with inconsistencies
at the quantum level (see for example \cite{Woodard}).

Recenty, one of us has proposed an action for Einstein gravity coupled with
a vector field that leads to modifications to conventional
gravitational interactions \cite{BG}. The action is constructed
strictly according to the rules for an effective field theory with a mass
scale cutoff, $M$: all Lorentz invariant terms containing the
vector field and metric up to a predefined order in $E/M$ (where $E$ is
the energy scale at which the theory is being considered) are included. The theory is manifestly self-consistent at energy scales below the cutoff $M$, 
and the usual problems of theories of modified gravity, such as ghosts \cite{Fierz1939}, strong-coupling \cite{Arkani-Hamed2002sp}, and discontinuities \cite{vanDam1970vg} are absent. The
vector field has a Lorentz-invariant mass term with the wrong sign, which leads to a vacuum
expectation value for the vector field and spontaneous breaking of Lorentz invariance. As a result one finds
that there are additional graviton modes and modified dispersion relations. Moreover, the new terms in the action lead to
additional terms in the equations of motion which affect the dynamics of the
universe on very large scales. It is this latter aspect of the theory that
we wish to explore in this paper.

Our work in this paper complements that presented in \cite{CL} where it
has been shown that Lorentz-violating vector fields may slow down the
expansion rate of the Universe. In their analysis, the authors consider
a fixed norm vector field, pinned down through a Lagrange multiplier term
in the action. Furthermore, they examine a simplification of
the theory presented in \cite{JM} by restricting themselves to quadratic
terms in the the vector field. In this paper we include higher
order terms and allow the norm of the vector field to vary. A fixed
norm case will be considered as a limiting case where a  coupling
constant in the potential for the vector field becomes infinite.

The layout of this paper is as follows. In Section \ref{recap} we
present the action and briefly discuss other attempts at studying
Lorentz violation from vector fields. In Section \ref{evol} we write down
the general equations of motion. We then specialise to the case of
a homogeneous and isotropic spacetime. As will become apparent, the
parameter space of this theory is immense. There are effectively
6 coupling constants and three energy scales that play a role. We briefly
discuss the range of parameters that need to be explored in \ref{para}.
In the following three sections we look at the dynamics in more detail.
In Section \ref{timelike} we focus on the case in which the norm of
the vector field is fixed. This corresponds to the case where a coupling
constant of the potential is infinite and the theory becomes
an extension of the theory studied in \cite{CL,JM}. In Section \ref{free}
we set the potential energy to zero and allow the vector field to roam freely.
This allows us to study the impact of the multiple derivative couplings
between the vector field and the metric. We then look at the more general
case in Section \ref{general} and conclude in Section \ref{conclusions}.

\section{An effective theory of spontaneously-broken
Lorentz invariance: a recap.}\label{recap}

The effective action for gravity coupled to a vector field can be written as
\begin{eqnarray}
S &=& S_{G}[\textbf{g},\textbf{A}] + S_{m}, \\
S_{G}[\textbf{g},\textbf{A}] &=& \int d^{4}x(-g)^{1/2}\Bigg(\frac{M^{2}_{PL}}{2}R
-\frac{1}{4g^{2}}F^{ab}F_{ab} \nonumber \\ &&
-\frac{\alpha_{1}}{2}R_{ab}A^{a}A^{b}
-\frac{\alpha_{2}}{2}(\nabla_{a}A^{a})^{2}
\nonumber\\ &&-\frac{\beta_{1}}{2M^{2}}F_{ab}F^{ac}A_{c}A^{b}
-
\frac{\beta_{2}}{2M^{2}}\nabla_{a}A^{a}\nabla_{b}A^{c}A^{b}A_{c}
\nonumber \\ &&
 -\frac{\beta_{3}}{2M^{2}}\nabla_{a}A^{b}\nabla_{c}A_{b}A^{a}A^{c}\nonumber \\ && -\frac{\beta_{4}}{2M^{2}}\nabla_{a}A^{b}\nabla_{c}A^{d}A^{a}A^{c}A_{b}A_{d}
\nonumber \\&&-\frac{\gamma}{2}(A_{a}A^{a}
-M^{2}n_{a}n^{a})^{2}+....\Bigg),
\\
S_{m} &=& \int d^{4}x (-g)^{\frac{1}{2}} \textit{L}_{m}.
\end{eqnarray}
The gauge coupling constant is small ($g\ll 1$) and the dimensionless coupling constants, ($\alpha_1$,$\alpha_2$,
$\beta_1$,$\beta_2$,$\beta_3$,$\beta_4$,$\gamma$)
and $n_an^a$ are of order unity. Note that $n_{a}n^{a}$ is to be fixed during variation. The ellipsis denotes terms suppressed by powers of $M$ or $M_{PL}$; note that such a theory leads to corrections
to General relativity of order $M^2/M^2_{PL}$ in the weak-field limit. The $S_m$ contains all other
fields that contribute to the total action.

We can rewrite the action $S_{G}$ in a form akin to that presented in \cite{JM} as
\begin{eqnarray}
S_{G}[\textbf{g},\textbf{A}] &=& \frac{1}{16\pi G}\int d^{4}x (-g)^{\frac{1}{2}}[R+K^{ab}_{\phantom{ab}mn}\nabla_{a}A^{m}\nabla_{b}A^{n} \nonumber\\
& & -\gamma 8\pi G(A^{a}A_{a}-M^2n^{a}n_{a})^{2}], \label{action2}
\end{eqnarray}
where the kinetic kernel is defined through
\begin{eqnarray}
K^{ab}_{\phantom{ab}mn} &=& K(\textsc{1})^{ab}_{\phantom{ab}mn}+K(\textsc{2})^{ab}_{\phantom{ab}mn}, \nonumber \\
K(1)^{ab}_{\phantom{ab}mn} &=& c_{1}g^{ab}g_{mn} + c_{2}\delta^{a}_{\phantom{a}m}\delta^{b}_{\phantom{b}n} \nonumber \\
\nonumber && +c_{3}\delta^{a}_{\phantom{a}n}\delta^{b}_{\phantom{b}m}+c_{4}A^{a}A^{b}g_{mn}, \nonumber \\
K(2)^{ab}_{\phantom{ab}mn} &=& c_{5}\delta^{a}_{\phantom{b}n}A^{b}A_{m}+c_{6}g^{ab}A_{m}A_{n} \nonumber \\
&& +c_{7}\delta^{a}_{\phantom{a}m}A^{b}A_{n} +c_{8}A^{a}A^{b}A_{m}A_{n}
\nonumber
\end{eqnarray}
and the dimensionful coupling constants, $c_i$ are related to the dimensionless
ones through
\begin{eqnarray}
c_{1} &=& -\frac{8\pi G}{g^{2}},\nonumber \\
c_{2} &=& -8\pi G(\alpha_{2}-\alpha_{1}),\nonumber \\
c_{3} &=& -8\pi G(\alpha_{1}-\frac{1}{g^{2}}),\nonumber \\
c_{4} &=& -\frac{8\pi G(\beta_{1}+\beta_{3})}{M^{2}},\nonumber \\
c_{5} &=& +\frac{16\pi G\beta_{1}}{M^{2}},\nonumber \\
c_{6} &=& -\frac{8\pi G\beta_{1}}{M^{2}},\nonumber \\
c_{7} &=& -\frac{8\pi G\beta_{2}}{M^{2}},\nonumber \\
c_{8} &=& -\frac{8\pi G\beta_{4}}{M^{4}}. \nonumber
\end{eqnarray}

As compared to the theories studied in \cite{CL,JM} we are not restricting the vector field to have fixed norm, nor are we restricting the vector field
to be exactly timelike.

\section{Cosmological Evolution Equations}\label{evol}
We can find the equations of motion for the effective theory in
a general form using the formulation of(\ref{action2}).
The Einstein equations are
\begin{equation}
G_{ab}=\tilde{T}_{ab}+8\pi GT^{m}_{ab}, \nonumber
\end{equation}
where $G_{ab}$ is the Einstein tensor and $T^{m}_{ab}$ is the
energy-momentum tensor of the fields contained in $S_m$.$\tilde{T}_{ab}$
is defined to be
\begin{eqnarray}
\tilde{T}_{ab}&\equiv&\frac{1}{2}\nabla_{m}\left[ I_{(a}^{\phantom{a}m}A_{b)}-I^{m}_{\phantom{m}(a}A_{b)}-I_{(ab)}A^{m}\right]\nonumber \\
\nonumber && + \frac{1}{2}\nabla_{m}\left[ J_{(a}^{\phantom{a}m}A_{b)}-J^{m}_{\phantom{m}(a}A_{b)}-J_{(ab)}A^{m}\right]\\
\nonumber && + c_{1}\left[ (\nabla_{m}A_{a})(\nabla^{m}A_{b})-(\nabla_{a}A_{m})(\nabla_{b}A^{m})\right]\\
\nonumber && + c_{4}A^{m}\nabla_{m}A_{a}A^{n}\nabla_{n}A_{b}\\
\nonumber && + c_{5} \delta^{m}_{\phantom{m}p}A^{n}\nabla_{n}A^{p}A_{(b|}\nabla_{m}A_{|a)}\\
\nonumber && + c_{6}\left[2\nabla^{m}A_{(a|}A_{n}A_{|b)}\nabla_{m}A^{n}-A_{m}A_{n}\nabla_{a}A^{m}\nabla_{b}A^{n}\right]\\
\nonumber && + c_{7} \delta^{m}_{\phantom{m}n}A^{p}\nabla_{p}A_{(a|}\nabla_{m}A^{n}A_{|b)}\\
\nonumber && + c_{8}2\nabla_{m}A_{(a|}\nabla_{n}A^{p}A_{|b)}A_{p}A^{m}A^{n}\\
\nonumber &&  + \frac{1}{2}g_{ab}K\\
&&-4\pi G\gamma\left[g_{ab}V+4A_{a}A_{b}\sqrt{V}\right],
\end{eqnarray}
in which
\begin{eqnarray}
I^{b}_{\phantom{b}n} &=& K^{ab}_{\phantom{ab}mn}\nabla_{a}A^{m}, \nonumber\\
J^{a}_{\phantom{a}m} &=& K^{ab}_{\phantom{ab}mn}\nabla_{b}A^{n},\nonumber\\
K &=& K^{ab}_{\phantom{ab}mn}\nabla_{a}A^{m}\nabla_{b}A^{n},\nonumber\\
\sqrt{V} &=& (g_{ab}A^{a}A^{b}-M^2n^{a}n_{a}) \label{tensorA}.
\end{eqnarray}

The vector field equation of motion is given by
\begin{eqnarray}
\label{eq:vector}
0 &=& \nabla_{a}I^{a}_{\phantom{a}m}+\nabla_{a}J^{a}_{\phantom{a}m}-2c_{4}A^{e}\nabla_{e}A_{a}\nabla_{m}A^{a} \nonumber \\
\nonumber &&-c_{5}\delta^{a}_{\phantom{a}n}A_{e}\nabla_{a}A^{e}\nabla_{m}A^{n}-c_{5}\delta^{a}_{\phantom{a}n}A^{b}g_{em}\nabla_{a}A^{m}\nabla_{b}A^{n}\\
\nonumber &&-2c_{6}g^{ab}A_{n}g_{em}\nabla_{a}A^{e}\nabla_{b}A^{n}\\
\nonumber && -c_{7}\delta^{a}_{\phantom{a}e}A_{n}\nabla_{a}A^{e}\nabla_{m}A^{n}-c_{7}g_{nm}\delta^{a}_{\phantom{a}e}A^{b}\nabla_{a}A^{e}\nabla_{b}A^{n}\\
\nonumber && -2c_{8}A^{b}A_{a}A_{n}\nabla_{m}A^{a}\nabla_{b}A^{n}\\
\nonumber && -2c_{8}g_{em}A^{a}A^{b}A_{n}\nabla_{a}A^{e}\nabla_{b}A^{n}\\
 && +32\pi G\gamma \sqrt{V}g_{ma}A^{a}. \label{evolA}
\end{eqnarray}

We now wish to restrict ourselves to a homogeneous and isotropic universe, i.e.
one in which the metric is of the form
\begin{equation}
\textbf{g}= -\textbf{dt}^{2}+a(t)^{2}\delta_{ij}\textbf{dx}^{i}\otimes\textbf{dx}^{j} \nonumber
\end{equation}
(where we have restricted ourselves to Euclidean spatial sections). To do
so we must pick $n_an^a=-1$, i.e. the minimum of the vector field must
be time-like(A spacelike minimum would require the vector field to have spatial components and would violate isotropy). 
The vector field then has the isotropic and homogeneous form
\begin{equation}
\textbf{A} = (A(t),0,0,0) \nonumber.
\end{equation}
The contribution from $S_m$ is taken to be:
\begin{equation}
\textbf{T}^{m} = \rho \textbf{U}\otimes\textbf{U} +P(\textbf{g} + \textbf{U}\otimes\textbf{U}), \nonumber
\end{equation}
where $g_{ab}U^{a}U^{b} = -1$.

The equations of motion now simplify dramatically. The non-vanishing Christoffel symbols are
\begin{eqnarray}
\Gamma^{0}_{11}=\Gamma^{0}_{22}=\Gamma^{0}_{33}=a\dot{a}, \nonumber\\
\Gamma^{1}_{01}=\Gamma^{1}_{10}=\Gamma^{2}_{02}=\Gamma^{2}_{20}=\Gamma^{3}_{30}=\Gamma^{3}_{03}=\frac{\dot{a}}{a}, \nonumber
\end{eqnarray}
which can be used to calculate various contributions to equations
\ref{tensorA} and \ref{evolA}, for example
\begin{eqnarray}
\nabla_{e}A^{e}&=&\dot{A}+3\frac{\dot{a}A}{a}, \nonumber \\
\sqrt{V} &=& -(A^{2}-M^{2}n^{0}n^{0}), \nonumber \\
K &=& (c_{1}+c_{3})(\dot{A}^{2}+3\frac{\dot{a}^{2}}{a^{2}}A^{2})+c_{2}(\nabla_{e}A^{e})^{2} \nonumber \\
\nonumber &&-(c_{4}+c_{5}+c_{6}-c_{8}A^{2})A^{2}\dot{A}^{2}\\
\nonumber && - c_{7}A^{2}\dot{A}\nabla_{e}A^{e},
\end{eqnarray}
and the dot denotes differentiation with respect to $t$.

The Friedmann equation is now modified. From the $00$th component of the
Einstein equations we have
\begin{eqnarray}
\frac{3\dot{a}^{2}}{a^{2}} &=& 8\pi G\rho + \sum c_{i}\Lambda_{i}
\nonumber \\&& +4\pi G\gamma(V-4\sqrt{V}A^{2}),
\end{eqnarray}
where
\begin{eqnarray}
\Lambda_{1} &=& -\frac{3}{2}\frac{\dot{a}^{2}}{a^{2}}A^{2}+3\frac{\dot{a}}{a}A\dot{A}+\frac{\dot{A}^{2}}{2}+A\ddot{A},\nonumber \\
\Lambda_{2} &=& +\frac{3}{2}\frac{\dot{a}^{2}}{a^{2}}A^{2}+ 6\frac{\dot{a}}{a}\dot{A}A+3\frac{\ddot{a}}{a}A^{2}+A\ddot{A}+\frac{\dot{A}^{2}}{2},\nonumber \\
\Lambda_{3} &=& -\frac{3}{2}\frac{\dot{a}^{2}}{a^{2}}A^{2}+3\frac{\dot{a}}{a}A\dot{A}+\frac{\dot{A}^{2}}{2}+A\ddot{A},\nonumber \\
\Lambda_{4} &=& -3\frac{\dot{a}}{a}A^{3}\dot{A}-\frac{3}{2}A^{2}\dot{A}^{2}-A^{3}\ddot{A},\nonumber \\
\Lambda_{5} &=& -3\frac{\dot{a}}{a}A^{3}\dot{A}-\frac{3}{2}A^{2}\dot{A}^{2}-A^{3}\ddot{A},\nonumber \\
\Lambda_{6} &=& -3\frac{\dot{a}}{a}A^{3}\dot{A}-\frac{3}{2}A^{2}\dot{A}^{2}-A^{3}\ddot{A},\nonumber \\
\Lambda_{7} &=& -3\frac{\dot{a}^{2}}{a^{2}}A^{4}-\frac{9}{2}\frac{\dot{a}}{a}A^{3}\dot{A}-\frac{3}{2}\frac{\ddot{a}}{a}A^{4}\nonumber \\
\nonumber && -\frac{3}{2}A^{2}\dot{A}^{2}-A^{3}\ddot{A},\nonumber \\
\Lambda_{8} &=& 3\frac{\dot{a}}{a}A^{5}\dot{A}+\frac{5}{2}A^{4}\dot{A}^{2}+A^{5}\ddot{A}. \nonumber
\end{eqnarray}
From the trace of the spatial part of the Einstein equations we find
\begin{eqnarray}
-2\frac{\ddot{a}}{a}-\frac{\dot{a}^{2}}{a^{2}} &=& 8\pi GP+\sum c_{i}\Upsilon_{i}\nonumber \\
\nonumber && -4\pi G\gamma V,
\end{eqnarray}
where
\begin{eqnarray}
\Upsilon_{1} &=& -\frac{1}{2}\frac{\dot{a}^{2}}{a^{2}}A^{2}-\frac{\ddot{a}}{a}A^{2}-2\frac{\dot{a}}{a}A\dot{A}+\frac{1}{2}\dot{A}^{2},\nonumber \\
\Upsilon_{2} &=& -3\frac{\ddot{a}}{a}A^{2}-\frac{3}{2}\frac{\dot{a}^{2}}{a^{2}}A^{2}-6\frac{\dot{a}}{a}A\dot{A}-\frac{1}{2}\dot{A}^{2}-A\ddot{A},\nonumber \\
\Upsilon_{3} &=&
-\frac{1}{2}\frac{\dot{a}^{2}}{a^{2}}A^{2}-\frac{\ddot{a}}{a}A^{2}-2\frac{\dot{a}}{a}A\dot{A}+\frac{1}{2}\dot{A}^{2},\nonumber \\
\Upsilon_{4} &=& -\frac{1}{2}A^{2}\dot{A}^{2},\nonumber \\
\Upsilon_{5} &=& -\frac{1}{2}A^{2}\dot{A}^{2},\nonumber \\
\Upsilon_{6} &=& -\frac{1}{2}A^{2}\dot{A}^{2},\nonumber \\
\Upsilon_{7} &=& \frac{1}{2}A^{2}\dot{A}^{2}+\frac{1}{2}A^{3}\ddot{A},\nonumber \\
\Upsilon_{8} &=& +\frac{1}{2}A^{4}\dot{A}^{2}.\nonumber
\end{eqnarray}
The equation of motion for the vector field becomes
\begin{equation}
0 = \sum c_{i}\Psi_{i}-16\pi G\gamma \sqrt{V}A,
\end{equation}
where
\begin{eqnarray}
\Psi_{1} &=& \ddot{A}+3\frac{\dot{a}}{a}\dot{A}-3\frac{\dot{a}^{2}}{a^{2}}A ,\nonumber \\
\Psi_{2} &=& 3\frac{\ddot{a}}{a}A-3\frac{\dot{a}^{2}}{a^{2}}A+3\frac{\dot{a}}{a}\dot{A}+\ddot{A},\nonumber \\
\Psi_{3} &=& \ddot{A}+3\frac{\dot{a}}{a}\dot{A}-3\frac{\dot{a}^{2}}{a^{2}}A,\nonumber \\
\Psi_{4} &=& -3\frac{\dot{a}}{a}A^{2}\dot{A}-A\dot{A}^{2}-A^{2}\ddot{A},\nonumber \\
\Psi_{5} &=& -3\frac{\dot{a}}{a}A^{2}\dot{A}-A\dot{A}^{2}-A^{2}\ddot{A},\nonumber \\
\Psi_{6} &=& -3\frac{\dot{a}}{a}A^{2}\dot{A}-A\dot{A}^{2}-A^{2}\ddot{A},\nonumber \\
\Psi_{7} &=& +\frac{3}{2}\frac{\ddot{a}}{a}A^{3}+3A^{3}\frac{\dot{a}^{2}}{a^{2}}+3\frac{\dot{a}}{a}A^{2}\dot{A}-A\dot{A}^{2}-A^{2}\ddot{A},\nonumber \\
\Psi_{8} &=& 3\frac{\dot{a}}{a}A^{4}\dot{A}+4A^{3}\dot{A}^{2}+A^{3}\ddot{A}. \nonumber
\end{eqnarray}

The additional terms significantly complicate the
equations of motion. In particular, we now find that the $00$ equation (a
constraint equation in general relativity) contains second derivatives of the scale factor
which appear in $\tilde{T}_{ab}$. This situation is similar to the situation one encounters in higher-derivative theories of
gravity. It is not unexpected: in \cite{BG}, two additional graviton modes are found,
a clear indication that we should find more degrees of freedom in this
theory than are present in conventional Einstein gravity.

\section{The Parameter Space}\label{para}
Because of the large number of independent, Lorentz-invariant, two-derivative terms in the effective action, 
the parameter space for this theory is large. Even though
we have restricted ourselves to a homogeneous and isotropic universe,
we must still contend with eight coefficients, $c_i$, and the
coupling constant, $\gamma$. Inspecting the
equations of motion, we find that there are degeneracies. $c_1$ and $c_3$
multiply equivalent terms as do $c_4$, $c_5$ and $c_6$. Putting them all
together and reverting back to dimensionless parameters we find that
we are left with six independent parameters. They can be organized into various
groups
\begin{description}
\item[Kinetic terms, quadratic in $\textbf{A}$:] $\alpha_1$ and $\alpha_2$
\item[Kinetic terms, quartic in $\textbf{A}$:] $\beta_2$ and $\beta_3$
\item[Kinetic terms, sextic in $\textbf{A}$:] $\beta_4$
\item[Potential energy terms:] $\gamma$
\end{description}
Unless an additional symmetry principle is proposed, none of these terms
can be discarded. In order to illustrate the possible dynamics, we shall restrict ourselves to
sub-spaces of the full parameter space in what follows; the full dynamics are simply too complex.
But we emphasize that
the true spirit of effective field theories does not allow us to selectively
discard terms.

There are also three relevant dimensionful scales, namely the Planck scale $M_P$, the cutoff $M$, and the observation scale $M_0$.
From these, we define the two dimensionless ratios $r = M/M_P$ and $r_0 = M_0/M_P$. Corrections to General Relativity in the weak field limit are of order $r^{2}$.
The observation scale can be defined in terms of the energy density, $\rho$
by fixing the scale factor, $a$ to be unity at the time of observation. We then
have
\begin{eqnarray}
\rho\equiv\frac{M_0^4}{a^n}, \nonumber
\end{eqnarray}
where $n$ depends on the equation of state of the matter component.
So, in addition to the six dimensionless parameters we have two dimensionless
energy scales.

We clearly have a very large space of parameters to explore. To do so, and with the aim of illustrating qualitatively the possible cosmological dynamics, 
we will restrict ourselves to considering subspaces of the full parameter space in what follows. To this end, in the 
rest of this paper, we will try to identify what kind
of behaviour each of the terms in the effective action, pinned to a given parameter, will generate.

\section{Strong coupling limit: $\gamma\rightarrow \infty$}\label{timelike}

In the limit of $\gamma$ tending to infinity, we expect the vector field to be fixed at the minimum of the
potential. This can also be realized by including a fixed-norm constraint to the action, of the form
\begin{equation}
\frac{1}{16\pi G}\int d^{4}x(-g)^{\frac{1}{2}} \lambda(A^{a}A_{a}-M^{2}n^{a}n_{a}), \nonumber
\end{equation}
where $\lambda$ is a Lagrange multiplier, which yields the constraint upon variation. Now, the time derivatives of \textbf{A} vanish. If $n^{a}n_{a}=-1$, as if the norm of a timelike unit vector, then we have \textbf{A}=(M,0,0,0).
The equations obtained by varying the action with respect to the vector field and the 00th component of the metric then reduce to
\begin{eqnarray}
0 &=& 3\frac{\dot{a}^{2}}{a^{2}}(c_{1}+c_{2}+c_{3}+c_{7}M^{2})\\
\nonumber && -3\frac{\ddot{a}}{a}(c_{2}-\frac{1}{2}c_{7}M^{2})-\lambda, \\
3\frac{\dot{a}^{2}}{a^{2}} &=& 8\pi G\rho -\frac{3}{2}\frac{\dot{a}^{2}}{a^{2}}M^{2}(c_{1}-c_{2}+c_{3}+2c_{7}M^{2})\\
\nonumber &&+3\frac{\ddot{a}}{a}M^{2}
(c_{2}-\frac{1}{2}c_{7}M^{2})+\lambda M^{2}.
\end{eqnarray}
Eliminating the Lagrange multiplier $\lambda$ and converting to $\alpha$ and $\beta$ coefficients yields the
following modified Friedmann equation
\begin{equation}
(1+(\frac{3}{2}\alpha_{2}-\alpha_{1})r^{2})3\frac{\dot{a}^{2}}{a^{2}}=8\pi G\rho.
\end{equation}
In this limit, we then have an effective rescaling of the gravitational 
constant $G$ in the cosmological background, as found in \cite{CL}.

\section{Weak coupling limit: $\gamma\rightarrow 0$}\label{free}

Let us now look at the dynamics of the system which result from discarding the
potential term. The vector field is now free to
vary subject to the couplings between its kinetic terms and the
metric. 

We first focus on the case in which $\alpha_1\neq 0$,
with the remaining coupling constants vanishing.
Varying with respect to the vector and $ii$th component of the metric yields
\begin{eqnarray}
3A\frac{\ddot a}{a}&=&0, \nonumber \\
2\frac{\ddot a}{a}+\left(\frac{\dot a}{a}\right)^2&=&
-8\pi G P+\alpha_1 8\pi G [A{\ddot A}+2A^2\frac{\ddot a}{a}
 \nonumber \\
&& +\left(\frac{\dot a}{a}\right)^2A^2+4\frac{\dot a}{a}A{\dot A}+{\dot A}^2].
\nonumber
\end{eqnarray}
The first of these equations has the solution
\begin{eqnarray}
a(t)=\frac{{\dot s}_ia_i}{t_i}
(t+t_i\frac{1-{\dot s}_i}{{\dot s}_i}), \nonumber
\end{eqnarray}
where $a_i$ and ${\dot s}_i$ are the intial conditions at
time $t_i$. Note that we can always rescale
$t\rightarrow t+t_i\frac{1-{\dot s}_i}{{\dot s}_i}$, such that $a\propto t$, and we
will do so from now on.

To solve the second
equation we define a new variable, $X=\frac{1}{2}a^4A^2$.
We then have
\begin{eqnarray}
{\dot X}&=&2a^3{\dot a}A^2+a^4A{\dot A}, \nonumber \\
{\ddot X}&=& a^4A{\ddot A}+2a^3{\ddot a}A^2+a^4{\dot A}^2
+6a^2{\dot a}^2A^2+8a^3{\dot a}{\dot A}A . \nonumber
\end{eqnarray}
We can now rewrite the second equation as
\begin{eqnarray}
{\ddot X}-4\frac{\dot a}{a}{\dot X}+6\left(\frac{\dot a}{a}\right)^2X=
\frac{1}{8\pi G\alpha_1}({\dot a}^2a^2+8\pi G Pa^4), \nonumber
\end{eqnarray}
whose homogeneous part has the solution
\begin{eqnarray}
X=At^2+Bt^3. \nonumber
\end{eqnarray}
The dominant term at late times comes from the particular integral
\begin{eqnarray}
X=\frac{({\dot s}_ia_i)^4}{16\pi G\alpha_1t_i^4}t^4. \nonumber
\end{eqnarray}
The particular integrals due to the pressure term of the
matter/radiation are
\begin{eqnarray}
X&=& \frac{P_0}{5\alpha_1}t^2\ln t, \ \ \ \mbox{radiation era}, \nonumber \\
X&=& \frac{P_0}{\alpha_1}t^3\ln t, \ \ \ \mbox{matter era}. \nonumber
\end{eqnarray}
We can find the vector field from the definition of $X$. The
dominant solution is
\begin{eqnarray}
A^2\simeq \frac{1}{2\alpha_1}M^{2}_{Pl} .\nonumber
\end{eqnarray}
As we can see, in this reduced space of parameters
we have an attractor solution
given by $a\propto t$ and $A=\sqrt{1/2\alpha_1}M_{Pl}$.

An altogether different type of behaviour emerges if we consider
all constants to be zero, except for $\alpha_2$.
The two equations to solve now are:
\begin{eqnarray}
{\ddot A}&+&3\frac{\ddot a}{a}A-3(\frac{\dot a}{a})^2A+
3\frac{\dot a}{a}{\dot A}=0 ,\nonumber \\
2\frac{\ddot a}{a}&+&\left(\frac{\dot a}{a}\right)^2=
-8\pi G P-8\pi G\alpha_2 [A{\ddot A}+3A^2\frac{\ddot a}{a}
+\frac{3}{2}(\frac{\dot a}{a})^2A^2 \nonumber \\
&&+6\frac{\dot a}{a}A{\dot A}
+\frac{1}{2}{\dot A}^2] .\nonumber
\end{eqnarray}
Defining $Y=a^3A$,we find that the first equation reduces to
\begin{eqnarray}
{\ddot Y}-3\frac{\dot a}{a}{\dot Y}=0 ,\nonumber
\end{eqnarray}
with solution
\begin{eqnarray}
{\dot Y}=Ca^3 ,\nonumber \\
Y=C\int dt a^3 +D .\nonumber
\end{eqnarray}
We can also rewrite the second equation in terms of $Y$, which
conveniently simplifies to
\begin{eqnarray}
2\frac{\ddot a}{a}+\left(\frac{\dot a}{a}\right)^2=
-8\pi G P-4\pi G\alpha_{2}\frac{{\dot Y}^2}{a^6} \nonumber
\end{eqnarray}

Let us consider the case where  $P$ is negligible. Then, with $b=\ln a$, we find
\begin{eqnarray}
{\ddot b}+ \frac{3}{2}{\dot b}^2= -2\pi G\alpha_{2}C^{2}\equiv J,
\nonumber
\end{eqnarray}
which can be solved as follows. Firstly, we consider separately the cases where $J$ is positive or negative, denoted by $J_{+}$ and $J_{-}$ respectively. Integrating, we find
\begin{eqnarray}
\dot{b} = \sqrt{\frac{2J_{+}}{3}}\tanh(\sqrt{\frac{3J_{+}}{2}}(t-t_{a}))
\nonumber
\end{eqnarray}
and
\begin{eqnarray}
\dot{b} = -\sqrt{\frac{-2J_{-}}{3}}\tan(\sqrt{\frac{-3J_{-}}{2}}(t-t_{a})),
\nonumber
\end{eqnarray}
respectively, where $t_{a}$ is a constant of integration. Integrating again yields
\begin{eqnarray}
b = \frac{2}{3}\ln(\cosh(\sqrt{\frac{3J_{+}}{2}}(t-t_{a})))+D
\nonumber
\end{eqnarray}
and
\begin{eqnarray}
b = \frac{2}{3}\ln(\cos(\sqrt{\frac{-J_{-}3}{2}}(t-t_{a})))+D,
\end{eqnarray}
where D is an integration constant. We now have a complete 
solution for $a=\exp(b)$. Note that
the solution is singular, with $a\rightarrow 0$ in a finite time, for $J<0$.

This type of behaviour arises because of the novel
type of coupling that arises when considering a vector field. The Lorentz 
structure of the vector field leads to a coupling with derivatives of
the metric and hence to second derivatives in the would-be constraint equations
and equations of motion. These lead to instabilities
in the solutions of the equations of motion, a finite time singularity in this case. These
instabilities presumably signal deeper pathologies in the effective field theory description.

If only $\beta_{3}$ is non-zero, the $00$-th and $ii$-th equations are, respectively,

\begin{eqnarray}
3(\frac{\dot{a}}{a})^{2} = 8\pi G\rho -\frac{8\pi G\beta_{3}}{M^{2}}(-3\frac{\dot{a}}{a}A^{3}\dot{A}-\frac{3}{2}A^{2}\dot{A}^{2}-A^{3}\ddot{A}),\nonumber\\
\nonumber
-2\frac{\ddot{a}}{a}-(\frac{\dot{a}}{a})^{2}=8\pi GP + \frac{8\pi G\beta_{3}}{M^{2}}(\frac{1}{2}A^{2}\dot{A}^{2}).
\end{eqnarray}
The vector field equation is
\begin{eqnarray}
0 = 3\frac{\dot{a}}{a}A^{3}\dot{A}+3A^{2}\dot{A}^{2}+A^{3}\ddot{A} .\nonumber
\end{eqnarray}
Multiplying the $ii$-th equation by three, and adding to the $00$-th equation we recover
\begin{eqnarray}
-6\frac{\ddot{a}}{a}=8\pi G(3P+\rho),
\nonumber
\end{eqnarray}
where we have used the vector equation. 
We see that, in this case, the evolution is identical to that of 
General Relativity. For the case of only $\beta_{4}\neq 0$, 
the same applies, with the caveat that, in the weak coupling limit, 
there is nothing to stop $A(t)$ approaching values beyond the
applicability of the effective action.

\begin{figure}[ht]
\epsfig{file=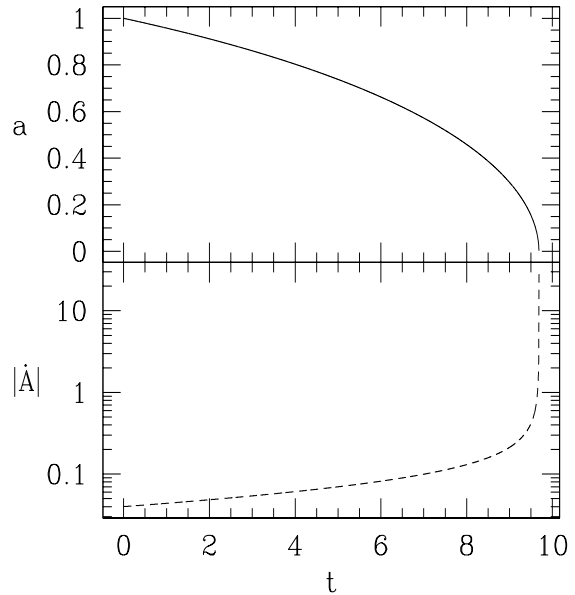,width=8.3cm,height=9.3cm}
\vskip -0.37in
\caption{
Weak coupling limit response when $\beta_{2}\neq 0$ and 
$\dot{A}(0)<0$ (note that the units are arbitrary).\label{Fig1}}
\end{figure}

Finally, we remark that the case in which $\beta_{2}\neq 0$ is significantly more complicated, given
the appearance of a wide variety of terms in each equation. We may, however,
combine the vector and $ii$-th equations so as to to obtain two equations, each 
containing just one type of second-order time derivative term.  In this way, we obtain expressions
for $\frac{\ddot{a}}{a}$ and $\ddot{A}$. It turns out that the $\ddot{A}$ equation
is singular at $A=0$ and $A=(8/3\beta_2)^{1/4}\sqrt{M M_P}$.
We will see later that in the corresponding case with a natural value of $\gamma$, physically viable solutions are obtained only for positive $\beta_{2}$. Moreover, for acceptable 
values of $A$ the second singularity will be real and lie above $A=M$. 
Solving the equations numerically, we find that for $\dot{A}(0)<0$ the system generically reaches the singularity
at $A=0$ (see Figure \ref{Fig1}), whilst for $\dot{A}(0)>0$ the system reaches the second singularity.

\section{The general case}\label{general}

We now consider solutions for natural values of the parameters $\alpha_{i},\beta_{i},\gamma$ (order unity). Furthermore, $\gamma$ is assumed to be positive. In \cite{BG}, it was argued that $r$ could be as high as $\sim 10^{-2}$ and as low as $\sim 10^{-31}$ and we will opt to look for behaviour between these bounds. Even with these considerations, the parameter space remains vast so we will proceed, as before, by considering the effect of isolated non-vanishing $\alpha_{i}$ and $\beta_{i}$ parameters. Following this, we will consider a few non-vanishing 
combinations of $\alpha_{i}$ and $\beta_{i}$ which yield novel behaviour. 

Reasonable initial conditions are 
\begin{eqnarray}
b(0) &=& 0 ,  \nonumber \\
\dot{b}(0)&\approx& \frac{rM}{3^{1/2}}, \nonumber \\ 
A(0) &=& M, \nonumber \\ 
|\dot{A}(0)|&\approx& rM^{2}. \nonumber
\end{eqnarray}
We will consider the universe to be radiation dominated.

\begin{figure}[ht]
\epsfig{file=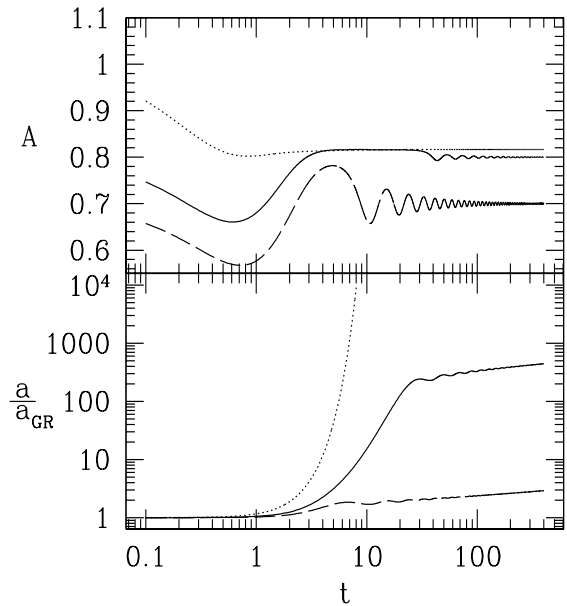,width=8.3cm,height=9.3cm}
\vskip -0.37in
\caption{
Response of the system for r=1 (dot), r=0.8 (solid),r=0.7 (dashed). $\alpha_{2}=-1$,$\gamma=1$ and
$M_{Pl}=1$.\label{Fig2}}
\end{figure}

Let us first consider {$\alpha_{1}\neq 0$} and all remaining constants
equal to zero. Recall that this is the case where only the
$R_{ab}A^{a}A^{b}$ term contributes in the action. From (\ref{eq:vector}) see that,
peculiarly, second derivatives of the vector field $\ddot{A}$ are
absent from the vector field equation and present in the metric field
equation whereas $\ddot{b}$ appears in both. Hence the
vector equation becomes an evolution equation for $b$

We find for all values of $r$ that the field $A$ undergoes damped oscillations, irrespective of the 
sign of $\alpha_{1}$. Indeed, 
for $r<<1$, $\ddot{A}$ 
is dominated by
\begin{eqnarray}
\ddot{A} = -\frac{4\gamma M(A^{2}-M^{2})}{3\alpha_{1}^{2}r^{2}A}+... ,\nonumber
\end{eqnarray}
where the ellipsis denotes damping terms, which are small independently of the sign of
$\alpha_{1}$. In the limit of $A \rightarrow M$, we have that 
$\ddot{b}\rightarrow 0$. Thus we generically approach a 
loitering solution where the scale factor $a$ grows linearly in time.

If we now consider only $\alpha_{2}\neq 0$, we note that second
derivatives of $b$ and $A$ appear in each field equation, an
indication of how the coupling to the vector field modifies the
kinetic terms of the metric. Consequently, for instance, terms in the
metric equation of order 1 or lower in time derivatives contribute to
$\ddot{A}$ to a degree suppressed by $r^{2}$. As in the weak coupling
limit \cite{BG}, the sign of $\alpha_{2}$ has an impact on the evolution.
It is found that positive values of $\alpha_{2}$ generically lead to
unbounded growth in $A$, whilst negative values (of order unity) lead
to oscillations about $M$, damped roughly on a timescale
$(rM)^{-1}$. Therefore, at times much greater than this, the vector
field will lie fixed at $A=M$, and we can read off the resulting
contribution to the metric equation as
$\tilde{T}_{ii}=(-3\alpha_{2}r^{2}/2)a^{-2}G_{ii}$. The evolution equation 
for the scale factor now becomes:
\begin{eqnarray}
\label{eq:a2track}
-2\frac{\ddot{a}}{a}-(\frac{\dot{a}}{a})^{2} = \frac{8\pi G}{1+\frac{3}{2}\alpha_{2}r^{2}}P .
\end{eqnarray}
We can interpret this as a rescaling of the gravitational constant
$G$. As discussed in \cite{CL}, this can only be detected by
comparison with regimes where the vector field may rescale $G$ by a
differing amount.

\begin{figure}[ht]
\epsfig{file=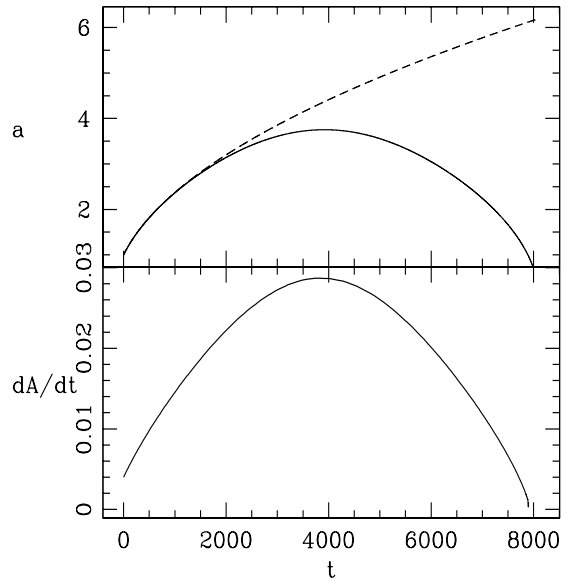,width=8.3cm,height=9.3cm}
\vskip -0.37in
\caption{
Top panel: Actual scale factor (solid) compared to 
GR scale factor (dashed evolution). 
Bottom panel: Oscillation amplitude of $\dot{\frac{A}{M}}$.
$r=0.02$, $b_{2}=1$,$\gamma=1$ and $M_{Pl}=10$\label{Fig3}}
\end{figure}

It is interesting to note that another consistent solution may emerge for
high values of $r$- that is values of order unity. In this range we see that $A$ may be attracted to a fixed solution at 
$A=\sqrt{\frac{2}{-3\alpha_{2}}}M_{Pl}\equiv B$, 
now close to M (see top panel of Figure \ref{Fig2}). At $A=B$ we have that
\begin{eqnarray}
\ddot{b} &=& -\frac{\gamma}{4M^{2}_{Pl}}(3B^{2}-M^{2})(B^{2}-M^{2}) ,\nonumber \\
\nonumber &\equiv & C.
\end{eqnarray} 
To understand the nature of this solution, we consider the evolution
of the ratio of the scale factor $a(t)$ to the scale factor $a_{GR}(t)$ resulting
from the same initial conditions but in the absence of the vector
field, i.e. as in pure general relativity. This ratio is plotted
in the bottom panel of Figure \ref{Fig2}.  The stability of this
solution depends on the sign of $C$: For $C>0$, $\dot{b}$
grows linearly with time and $A$ is constant, leading to an
asymptotic solution at late times of the form
\begin{eqnarray}
a(t) &\propto& e^{C(t-t_{a})^{2}},\\
A(t) &=& B, \nonumber
\end{eqnarray}
where $t_{a}$ is a constant of integration. 
In this limit the evolution of the vector field and metric are overwhelmingly
dominated by terms in $\dot{b}$. We can consider the stability of this solution by considering the 
evolution of small perturbations to $b$ and $A$. 
The first order perturbation $A_{1}$ to $A$ is found to obey
\begin{eqnarray}
\label{eq:pert}
\ddot{A_{1}}+3Ct\dot{A_{1}}+9C^{2}t^{2}A_{1}=0,
\end{eqnarray}

The general solution to this equation is:
\begin{eqnarray}
A_{1}(t) &=& k_{1} \frac{e^{\frac{-3Ct^{2}}{4}}}{t^{\frac{1}{2}}}
W_{\textit{M}}(\frac{i\sqrt{3}}{4},\frac{1}{4},
\frac{i\sqrt{3}Ct^{2}}{2}),\nonumber \\
&& +k_{2}\frac{e^{\frac{-3Ct^{2}}{4}}}{t^{\frac{1}{2}}}
W_{\textit{W}}(\frac{i\sqrt{3}}{4},\frac{1}{4},
\frac{i\sqrt{3}Ct^{2}}{2}),
\end{eqnarray}
where $k_{1}$ and $k_{2}$ are integration constants
and $W_{M}$ and $W_{W}$ are Whittaker $M$ and $W$ functions. 
We have checked that $A_{1}\rightarrow 0$ as $t\rightarrow \infty$ for the appropriate values of $r$, such that 
the solution is stable.

In the case that $C<0$, $\dot{b}$ will decrease linearly in time whilst the vector field is in the vicinity of $B$. Note that with the opposite sign of $C$, perturbations (as in (\ref{eq:pert})) are no longer stable. Thus we may expect that if the system can reach a fixed solution at $B$ it will only do so briefly before moving away. This is amply illustrated in figure \ref{Fig2}. For $r=1$ ($C>0$) the system quickly settles to the solution $A=B$ prompting runaway growth in $a/a_{GR}$. For $r=0.8$ ($C<0$) the system settles briefly at $A=B$, during which time $a/a_{GR}$ is enhanced significantly, before departing to the solution $A=M$. For $r=0.6$ (and all lower values of $r$) the vector field never reaches $B$ and settles to the $A=M$ tracking solution described by (\ref{eq:a2track}).

We now restrict ourselves to $\beta_{2}\neq 0$ and discard all other
terms.  We note that for the ranges of $r$ considered, the values at
which $\ddot{A}$ is singular (see \ref{free}) arguably lie beyond the
regime of validity of the effective action.  For negative $\beta_{2}$, $A$ experiences
unbounded growth in the direction of the initial perturbation, as in
Section \ref{free}, and so will generically reach singularities. For positive
$\beta_{2}$, the vector field undergoes oscillations about
$M$. Numerical exploration suggests that the system will evolve so
that the oscillations will be of increasing frequency and amplitude up
 until $t \approx (Mr)^{-1}$. Terms
 in $\tilde{\textbf{T}}$ then act to halt the expansion of the
universe leading to eventual collapse (see Figure \ref{Fig3}).

 If we consider $\beta_{3}\neq 0$ or $\beta_{4}\neq 0$, we
recover familiar behaviour. The vector field's stress energy tensor
contributes only first derivatives in time and terms such as
$\frac{\dot{a}}{a}^{2}$ and $\frac{\ddot{a}}{a}$ do not appear in the
vector field equation; thus, $A$ contributes and evolves much like a
scalar field. It is found that for positive values of $\beta_{3}$ and
for negative values of $\beta_{4}$, the vector field oscillates about
$A$. It undergoes Hubble friction, the oscillations decaying on a
timescale $\sim M^{-1}$. Settling towards $A=M$, the potential terms
vanish as before and so the effect on the expansion of the universe is
as in the weak-coupling limit i.e. identical to that of
general relativity. 

\begin{figure}[ht]
\epsfig{file=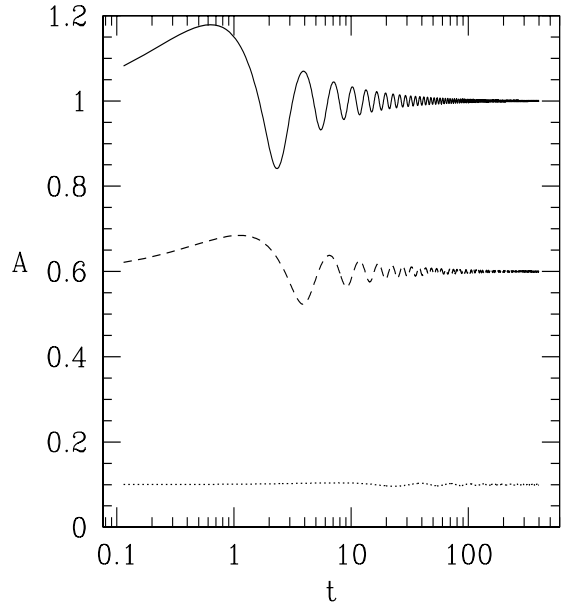,width=8.3cm,height=9.3cm}
\vskip -0.37in
\caption{
Comparative evolution of $A$ for $r$ values in ration 10:6:1 (solid:dashed:dotted) with $r=1$ being the highest scale. Such a high value is chosen only for ease of illustration. $\beta_{4}=-1$,$\gamma=1$, and $M_{pl}$=1. \label{Fig4}}
\end{figure}
w

An interesting simplification is obtained for particular the combination of coefficients, given by 
$\alpha_{1}=\alpha_{2}<0$.  With this combination, $\ddot{b}$
drops out from the vector equation. For acceptable values of $r$, the
vector field oscillates about $M$, the background expansion gradually
settling the field to this value. This happens on a timescale of
roughly $M^{-1}$. Thus at times larger than this the metric equation
reduces to
\begin{eqnarray}
-2\frac{\ddot{a}}{a}-(\frac{\dot{a}}{a})^{2} = \frac{8\pi G}{1-2r^{2}\alpha_{1}}P. \nonumber
\end{eqnarray}
Again this behaviour can be interpreted as a rescaling of $G$ or, in
this case, as the tensor $\tilde{\textbf{T}}$ tracking any form of
matter field in the universe.

Finally, we would like to make a general point about the appearance of singularities
in this system. The field equations can be written schematically as:
\begin{eqnarray}
\ddot{b}=\frac{1}{-f}(...), \nonumber \\
\ddot{A}=\frac{1}{f}(...), \nonumber
\end{eqnarray}
where $f$ is of the form
 \begin{eqnarray}
f&=&
(-2+r^{2}(2\alpha_{1}-3\alpha_{2}){\bar A}^{2})\\
\nonumber &&\times(-\alpha_{2}+(\beta_{3}+\beta_{2}){\bar A}^{2}-\beta_{4}{\bar A}^{4})\\
\nonumber &&-(\alpha_{1}-\alpha_{2}+\frac{\beta_{2}{\bar A}^{3}}{2})\\
\nonumber && \times(3\alpha_{1}-3\alpha_{2}-\frac{3\beta_{2}}{2}{\bar A}^{2})r^{2}{\bar A}^{2},
\end{eqnarray}
and where ${\bar A}=\frac{A}{M}$.
 When the terms in parentheses are nonvanishing, 
there will may occur singularities in the evolution for particular values of
${\bar A}$ . An example of this was encountered 
in \ref{free} for only $\beta_{2}\neq 0$.

We expect $r$ to be of order $10^{-2}$ or smaller and $\bar{A}$ to be
of order unity. Hence, generally, $(-2+r^{2}(2\alpha_{1}-3\alpha_{2})\bar{A}^{2})\approx -2$. Therefore, the second 
terms in parenthesis may be expected to be suppressed relative to the first by $r^{2}$. It is unlikely then that $f$ may vanish due to equality of the first and second terms in parenthesis. An alternative is for both groups of terms to vanish identically. This would seem to require of the first that $\delta \equiv-\alpha_{2}+(\beta_{3}+\beta_{2})\bar{A}^{2}-\beta_{4}\bar{A}^{4}$ vanishes along with at least one term from the second group. When considering isolated non-vanishing coefficients,  restrictions on their sign were found to be $\alpha_{2},\beta_{4}\leq 0$ and $\beta_{2},\beta_{3}\geq 0$ respectively. This would seem to imply that $\delta$ is inherently positive. It may only vanish when each of the coefficients are zero, in which case $f$ may only vanish if $\alpha_{1}$ is alzo zero. However, we emphasize that the restrictions on the sign of individual coefficients need not hold when general combinations of $\alpha_{i},\beta_{i}$ are considered.

\section{Conclusions}\label{conclusions}

In this paper, we have studied the cosmology of the model proposed in
\cite{BG}. As expected, there is a wide range of possible behaviours.
We achieve accelerated expansion if we have a term with $\alpha_2\neq0$
and a high mass scale $M$. The system will scale when $\alpha_1=\alpha_2$, or
when  $\alpha_2\neq0$ with a low mass scale. If $\beta_2\neq 0$, 
the universe will recollapse, while it will loiter if we only have 
$\alpha_1\neq0$. Clearly we have only looked at a small subset of the
parameter space but our analysis has allowed us to probe these different regimes. It also
allows us to make some comments about the structure of the theory.

Firstly, we stress again that this is an effective field theory, valid at energies below the
cutoff, $M$. In some cases, we have found that this theory is unstable, in the sense that
runaway solutions push the theory beyond its regime of validity. To properly
understand its behaviour in these regimes, we would need to at least consider higher order corrections, and ultimately some
ultra-violet completion of the theory would be necessary. Such a completion might conceivably simply involve extra fields
\cite{SM}, or, more likely, more complicated dynamics.

Secondly, the Lorentz structure of the vector field in this framework introduces
a novel phenomenon. For theories with $c_2\neq 0$ and $c_7\neq 0$ we find
second derivatives of the scale factor appear in what would have been (in standard General Relativity) constraint equations,
and in the evolution equations for the vector field. We find that this
leads to possible instabilities in the cosmology. Clearly, there is a
need for a complete perturbative analysis of this theory in a cosmological
setting, akin the study of Gauss-Bonnet or higher-derivative modifications to
gravity
 \cite{ghosts}.

Thirdly, the results found here can be added to known cosmological consequences of Lorentz 
violating fields: de Sitter expansion and dust-like stress energy tensor in background 
\cite{Arkani}, pervasive tracking solutions in background \cite{CL}, and 
generating the instability permitting the growth of large scale structure
\cite{DODEL}.

Finally, the theory discussed here, though complex in its structure, should be amenable to a detailed
comparison with current cosmological observations. We have laid down 
the framework for looking at constraints on the background evolution. The next
step is to construct the evolution equations for linear perturbations. With
these in hand, it should be possible to harness the wealth of new, high
precision cosmological observations and use them to provide constraints
on this theory.

{\it Acknowledgments}: The authors thank J. March-Russell, C. Skordis and G. Starkman for 
useful comments and discussions.


\end{document}